%% file: TQTN_arXive.tex
\documentclass[graybox,envcountchap]{TQTN_arXive}

\usepackage{mathptmx}
\usepackage{helvet}
\usepackage{courier}
\usepackage{type1cm}         

\usepackage{makeidx}         
\usepackage{imakeidx}
\usepackage{graphicx}        
\usepackage{multicol}        
\usepackage[bottom]{footmisc}

\makeindex             

\usepackage{amssymb,amsmath}

\usepackage{bm}

\usepackage{epsfig}
\usepackage{cite}
\usepackage[normalem]{ulem}

\begin{document}

\author{Dmitry A. Ryndyk}

\title{Theory of Quantum Transport \\ at Nanoscale}
\subtitle{introduction and contents \\ \null author version}

\date{The first edition is published as \\ {\it Springer Series in Solid-State Sciences 184}\\ (2016)}

\maketitle

\include{TQTN_title}
\include{TQTN_preface}

\include{TQTN_tableofcontents}
\include{TQTN_Intro}
\include{TQTN_Intro_Notations}
\include{TQTN_part_I}
\include{TQTN_LB}
\include{TQTN_GF}
\include{TQTN_TUN}
\include{TQTN_CB}
\include{TQTN_VP}
\include{TQTN_part_II}
\include{TQTN_NGF}
\include{TQTN_MW}
\include{TQTN_MB}
\include{TQTN_references}
\include{TQTN_printindex}

\bibliography{}

\end{document}

%% file: TQTN_preface.tex
\preface

This book is an introduction to a rapidly developing field of modern theoretical physics -- the theory of quantum transport at nanoscale. The theoretical methods considered in the book are in the basis of our understanding of charge, spin and heat transport in nanostructures and nanostructured materials and are widely used in nanoelectronics, molecular electronics, spin-dependent electronics (spintronics) and bio-electronics. Although some of these methods are already 20-25 years old, it is not so easy to find their systematic and consecutive description in one place. The main theoretical models and methods are distributed among many original publications, often written in different style and with different notations. During my research and teaching activities I had to overcome many obstacles to find required information. The results of my search I now present to your attention in more or less ordered form together with some original investigations.

The book is based on the lecture notes I used in the courses for graduate and post-graduate students the University of Regensburg and Technische Universit\"at Dresden (TU Dresden). Some basis knowledge of theoretical physics, especially quantum mechanics, is assumed. But otherwise I tried to make the text self-consistent and derive all formulas. In some cases I give the references to additional reading.

Since this book grew up from the lecture notes, I hope it will serve as the advanced-level textbook for Master and PhD students, and can be also interesting to the experts working in the fields of quantum transport theory and nanoscience. To this end I tried to combine modern theoretical results with the pedagogical level of explanations.  This book will help to cover, to some extend, the gap between undergraduate level textbooks and present day theoretical papers and reviews.

The book includes the Introduction and two parts. The first part is devoted to the basic concepts of quantum transport: Landauer-B\"uttiker method and matrix Green function formalism for coherent transport, Tunneling (Transfer) Hamiltonian and master equation methods for tunneling, Coulomb blockade, vibrons and polarons. The results in this part are obtained as possible without sophisticated techniques, such as nonequilibrium Green functions, which are considered in detail in the second part. We give a general introduction into the nonequilibrium Green function theory. We describe in detail the approach based on the equation-of-motion technique, as well as more sophisticated one based on the Dyson-Keldysh diagrammatic technique. The main attention is paid to the theoretical methods able to describe the nonequilibrium (at finite voltage) electron transport through interacting nanosystems, specifically the correlation effects due to electron-electron and electron-vibron interactions. We consider different levels of theoretical treatment, starting from a few-level model approach, such as a single-level electron-vibron (polaron) model and Hubbard-Anderson model for Coulomb interaction. The general formalism for multilevel systems is considered, and some important approximations are derived.

The book is focused on the ideas and techniques of quantum transport theory in discrete-level systems, as it is discussed in more detail in the Introduction. We do not consider here explicitly particular applications of the theory to semiconductor devices or the molecular transport theory based on atomistic, in particular density functional theory (DFT), methods.   

The Introduction includes a comprehensive literature review, but for sure not full. In the main text I did not try to cite all relevant publications, because it is more a textbook than a review. Nevertheless, I want to apologize for possibly missed important references.   

Despite of large efforts and time spent to improve the manuscript and to check it for misprints, the book definitely includes some mistakes, misprints, something is missed and should be added, as well as something would be better to remove. I'll greatly appreciate any comments and suggestions for improvement. 

\vspace{\baselineskip}
\begin{flushright}\noindent
Dresden,\hfill {\it Dmitry Ryndyk}\\
July 2015 \hfill { } \\
\end{flushright}

\extrachap{Acknowledgments}

First of all, I am grateful to Joachim Keller, Gianaurelio Cuniberti and Klaus Richter for their interest to my work and support during the years of my stay at the University of Regensburg and the Technische Universit\"at Dresden (TU Dresden). 

\null
I want to thank the members of the Nanoscale Modeling Group and the Chair for Materials Science and Nanotechnology at TU Dresden, especially Artem Fediai, Thomas Lehmann and Seddigheh Nikipar for many useful discussions and help with proof reading.

\null
Some of the results presented in this book were obtained in collaboration with Joachim Keller, Gianaurelio Cuniberti, Klaus Richter, Milena Grifoni, Andrea Donarini, Rafael Gutierrez, Bo Song, Michael Hartung, Pino D'Amico, Artem Fediai, Thomas Lehmann. I thank them and all other my colleagues in Regensburg and Dresden.

\null
Finally, I am thankful to the patience of my family.

%% file: TQTN_tableofcontents.tex
\tableofcontents 

%% file: TQTN_Intro.tex
\chapter{Introduction}

\section{Quantum transport in mesoscopic and nanoscale systems}
\index{Nanoscale systems}
\index{Mesoscopic systems}

What systems, models and methods are considered in this book? What is the meaning of the term ``nanoscale'' and what is its difference from the other known term ``mesoscopic''? 

One can note that {\it nanoscale} simply assumes {\it nanometer scale spatial dimensions}, very often any structure with at least one spatial dimension smaller than 100 nanometer (1\,nm = 10$^{-9}$ m) is considered as a subject of nanoscience. This definition, however, includes all types of nanostructures independently of their behavior and physical properties, which can be more or less quantum or even (semi)classical in the sense of required physical models. Many nanostructures actually  can be described by well established classical or semiclassical models.

We will focus on {\it quantum transport} of charge, spin and heat. {\it Nanoscale} in this respect characterizes not the size, but rather a specific type of systems and effects, which can be distinguished from both classical systems  and mesoscopic quantum systems.

If you insert the word ``nanoscale'' into the search line of your internet browser, you will probably find about 10 times more links than for the word ``mesoscopic''. Nevertheless, about 20 years ago, when the first books about quantum transport in mesoscopic systems and nanostructures had been published~\cite{Datta95book,Haug96book,Imry97book,Ferry97book,Dittrich98book}, almost any quantum transport was considered as  mesoscopic. Actually the term ``mesoscopic'' characterized the intermediate size between atomic (microscopic) and bulk (macroscopic). On the other hand, the main methods required to describe experiments in 80th years of 20th century, first of all the experiments with semiconductor hetero\-structures with $\mu$m sizes in transport directions, were based on the quasiclassical methods for quantum systems with dense (or even continuous) energy spectrum. Besides, the theoretical models were not based on discrete-level models and could be considered in language of real-space propagation paths and phase shifts. As a result, at present time, mesoscopic is associated with quasiclassical systems with continuous or dense energy spectrum.

But in last years, due to development of molecular electronics and computational methods for direct modeling at the atomic level, the methods specific for discrete-level systems become more and more important. At present, the nanoscale transport constitutes its own field of research separated not by hard walls, but by some visible boundaries from the rest large field of quantum mesoscopic transport. Let's try to estimate the parameters responsible for this boundary.

To some extend, the classification can be given based on the characteristic lengths and times. The most important scales are: \\
\indent $L$ -- the size of the system or characteristic internal length {\it in transport direction}; \\
\indent $l_p$, $\tau_p$ -- the elastic scattering length (mean free path) and time; \\
\indent $l_\epsilon$, $\tau_\epsilon$ -- the inelastic scattering (energy relaxation) length and time; \\
\indent $l_\varphi$, $\tau_\varphi$ -- the phase-decoherence length and time; \\
\indent $\lambda_B$ -- the de Broglie wave length (depends on the kinetic energy, for electrons in metals is taken at the Fermi surface).

\noindent Typically the characteristic lengths go in the following order  
$$\lambda_B<l_p<l_\phi<l_\epsilon.$$
For example, in semiconductor (GaAs,Si) 2D electron gas at low temperatures the values can be $\lambda_F\approx 0.05\ \mu m=50\ nm$, $l_p\approx 0.5\ \mu m$, $l_\varphi\approx 1\ \mu m$, $l_\epsilon\approx 3 \ \mu m$. In metals the numbers are similar: $l_\varphi\approx 1\ \mu m$ in gold at $T=1$ K. At room temperatures all these lengths in metals and semiconductors are very small and transport is described by semiclassical models. Note that it is not the case for carbon nanostructures like nanotubes, where even at room temperature both electron and phonon transport can be quantum.

Two scales from this list: the de Broglie wave length $\lambda_B$ and the phase-decoherence length $l_\varphi$ are specific for quantum transport (other exist also in the classical limit) and are most important for classification of transport regimes. In the case
$$\lambda_B\ll L\leqslant l_\varphi$$
the motion of electrons is phase-coherent and can not be described by classical equations, but in most cases is still quasiclassical, which means that classical trajectories can be used as a starting point and quantum effects are included mainly into the phases of quasiclassical wave functions. This is just a case of {\it mesoscopic} system.

Based on the definition of mesoscopic systems as the systems with continuous energy spectrum, we define {\it nanoscale} systems as the systems with essentially discrete energy spectrum in some parts. Usually it means that a discrete-level system is coupled to infinitely large electrodes (or substrate) with continuous spectrum.  For example, assume that the characteristic size of the central region in {\it transport direction} starts to be comparable with the electron wave length:
$$L\sim\lambda_B.$$
In this case quantization of the single-particle energy levels starts to be important. 

Of course, we are interested also in other cases when some system is naturally represented by discrete-level models. In particular, molecular junctions are described using the basis of atomic or molecular orbitals. One more origin of discrete {\it many-body} energy spectra is Coulomb interaction (the charging energy) in quantum dots and small grains. Finally, nanostructured low-dimensional materials (e.g. short nano\-tubes, graphene flakes, etc.) are described by discrete tight-binding (lattice) mo\-dels.

Thus, we suggest a point of view that the boundary between mesoscopic and nanoscale systems is mainly the boundary between: (i) continuous energy spectrum and continuous in real space equation for wave functions in mesoscopic case; and (ii) discrete energy spectrum and discrete basis wave functions in nanosystems. Of course, there is no strict separation between meso- and nano- transport and very often people actually mix these two terms. However, to have practical limits in the extremely wide field of nanoscience, I consider in this book only transport through {\it quantum nanosystems with discrete energy spectrum}, such as metal grains, semiconductor quantum dots and single molecules, coupled to one, two, or larger number of electrodes. 

We {\it do not consider} in this book the methods and approaches, which are typical only for mesoscopic transport and focus instead on specific for nanoscale transport questions. In particular, the following topics are not included: \\
\indent - quantum interference of Aharonov-Bohm type; \\
\indent - weak localization; \\
\indent - universal conductance fluctuations; \\
\indent - random matrix theory; \\
\indent - quantum Hall effect; \\
\indent - quasiclassical and semiclassical transport. \\
\noindent I refer the readers to numerous special reviews on mesoscopic transport.

Still, there are some topics important for both quasiclassical (mesoscopic) and pure quantum (nanoscale) systems, for example the Landauer scattering approach. That is the reason why we start from the ``mesoscopic'' Landauer-B\"{u}ttiker method in chapter \ref{LB}. However, in the next chapter \ref{chapter:GF} we formulate the Landauer approach for discrete basis using the technique of matrix Green functions, in such a way we get a nanoscale version of this approach.

There is one other significant peculiarity of nanoscale systems: the enhanced role of interactions. The theory of mesoscopic transport is based usually on free particles or weakly interacting particles, the perturbation theory is widely used. At nanoscale, as we already mentioned, both electron-electron and electron-vibron interactions may be strong and the 
Landauer approach can not be used anymore. Fortunately, we can use the powerful method of Nonequilibrium Green Functions, able to treat the many-body problems.

\begin{figure}[t]
\begin{center}
\epsfxsize=0.7\textwidth
\epsfbox{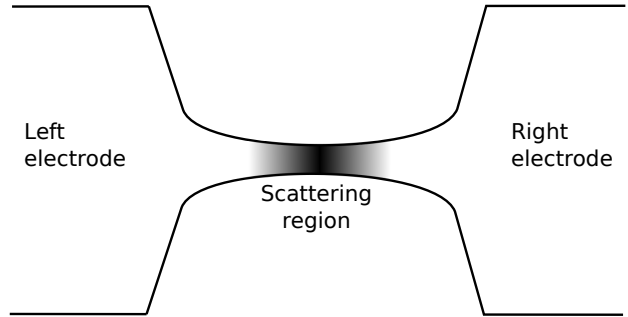}
\caption{Schematic picture of a nanojunction with strong coupling to the electrodes.}
\label{Intro_strong_coupling}
\end{center}
\end{figure}

\begin{figure}[b]
\begin{center}
\epsfxsize=0.7\textwidth
\epsfbox{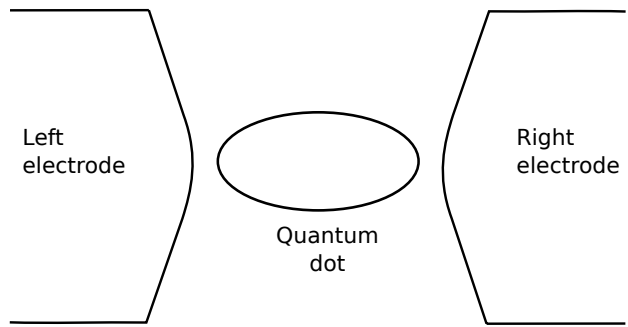}
\caption{Schematic picture of a nanojunction (quantum dot) with weak coupling to the electrodes.}
\label{Intro_weak_coupling}
\end{center}
\end{figure}

\begin{figure}[t]
\begin{center}
\epsfxsize=0.7\textwidth
\epsfbox{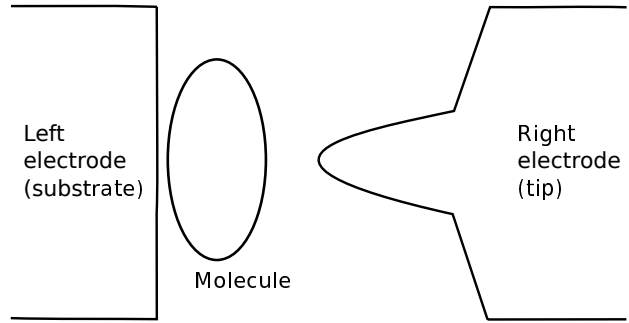}
\caption{Schematic picture of a strongly asymmetric nanojunction (STM set-up).}
\label{Intro_STM}
\end{center}
\end{figure}

\section{Nanojunctions}
\index{Nanojunction}

We focus on the models describing some {\it central system}, placed between two or many {\it ideal electrodes}, which assumed to be noninteracting and being in thermal equilibrium. On the contrary, the central system can be interacting and can be nonequilibrium if the finite voltage is applied. One can call such systems {\it nanojunctions}. Depending on the ratio between the energy scales associated with electron-electron or electron-vibron {\it interactions} in the central system (the examples of these energy scales are the effective charging energy and the polaron energy) and coupling to the leads, nanojunctions can be classified in several groups. 

In the case of strong coupling to the electrodes and weak interactions, the electronic states of the central system are hybridized with states in the electrodes, charge quantization is suppressed, transport is mainly coherent and the conductance is of the order of the conductance quantum $G_0=2e^2/h$. In some cases one can ignore completely the atomistic structure and formulate the model in the continuum medium approximation (a typical example is nanojunction shown in Fig.\,\ref{Intro_strong_coupling}), or use the lattice (tight-binding) model with given parameters. The basic way to understand quantum coherent transport in noninteracting systems is Landauer-B\"{u}ttiker method (usually formulated for atomistic or lattice systems with Green function formalism). We consider coherent transport in chapters \ref{LB} and \ref{chapter:GF}.

In the case of very weak coupling to the electrodes (Fig.\,\ref{Intro_weak_coupling}), the electronic states of the central system are only weakly disturbed, strong charge quantization and Coulomb blockade take place and transport is mainly determined by sequential tunneling. The central region in this case is often called {\it quantum dot}. In this case the master equation for probabilities of the many-body states is a good starting point. We consider different examples of sequential tunneling through the systems with Coulomb blockade and polaron effects in chapters \ref{chapter:CB} and \ref{VP}.

Besides, the important limiting case is a strongly asymmetric nanojunction (Fig.\,\ref{Intro_STM}), when the central region is strongly coupled to one electrode and weakly coupled to other one. This is a typical situation for STM experiments. The peculiarity of this case is that the central region (quantum dot, molecule) is in equilibrium or weakly nonequilibrium state even at large voltage, because it keeps the state in equilibrium with stronger coupled electrode. This type of junctions (as well as any direct contacts between two electrodes without any central region) can be describe by so-called Tunneling (or Transfer) Hamiltonian method without use of more sophisticated methods. We consider tunneling in chapter \ref{TUN}.

\section{From basic concepts to advanced methods}

The theoretical treatment of transport at the nanoscale (see introduction in~\cite{Datta95book,Haug96book,Imry97book,Ferry97book,Dittrich98book,Bruus04book,Cuniberti05book,Datta05book,DiVentra08book,Nazarov09book,Ferry09book,Cuevas10book}) requires the combined use of different techniques and approximations. We will consider discrete-level models starting from few-level and tight-binding noninteracting models and going in the direction towards the many-body models with strong electron-electron and electron-vibron interactions. Let us now outline the main concepts.

\index{Landauer-B\"{u}ttiker method !}
{\bf Landauer-B\"{u}ttiker method}~\cite{Landauer57ibm,Landauer70philmag,Economou81prl,FisherLee81prb,Buttiker85prb,Buttiker86prl,Landauer88ibm,Buttiker88ibm,Stone88ibm,Baranger89prb} establishes the fundamental relation between the wave functions (scattering amplitudes) of a junction and its conducting properties. The method can be applied to find the current through a noninteracting system or through an {\em effectively noninteracting} system, for example if the mean-field description is valid and the inelastic scattering is not essential. Such type of an electron transport is called coherent, because there is no phase-breaking and quantum interference is preserved during the electron motion across the system. In fact, coherence is assumed in many {\em ab initio} based transport methods based on density-functional theory and Landauer approach (DFT/Landauer), so that the Landauer-B\"{u}ttiker method is now routinely applied to any basic transport calculation through nanosystems and single molecules. Besides, it is directly applicable in many semiconductor quantum dot systems with weak electron-electron interactions. Due to simplicity and generality of this method, it is now widely accepted and is in the base of our understanding of coherent transport.

However, the peculiarity of single-molecule transport is just essential role of electron-electron and electron-vibron interactions, so that Landauer-B\"{u}ttiker method is not enough usually to describe essential physics even qualitatively.

The methods required to describe transport in weakly coupled junctions are usually different from the strong coupling case, because the effects of interactions are controlled by parameters $U/\Gamma$ for Coulomb interaction and $\lambda/\Gamma$ for electron-vibron interaction and become larger for tunneling junctions. Here $U$ is the characteristic Coulomb energy of electron-electron interaction (``Hubbard $U$''), $\lambda$ is the electron-vibron interaction constant, and $\Gamma$ is the coupling to electrodes.

During last years many new methods were developed to describe transport at finite voltage, with focus on correlation and inelastic effects, in particular in the cases when Coulomb blockade, Kondo effect and vibronic effects take place. There are two main theoretical frameworks that can be used to study quantum transport with interactions and at finite voltage: quantum master equation and nonequilibrium Green function techniques .

\index{Quantum Master Equation}
\index{Master equation ! quantum master equation}
{\bf Quantum Master Equation (QME)}\cite{Weiss99book,Breuer02book} is usually formulated in the basis of the many-body eigenstates of the molecule. It gives a fairly complete description of sequential tunneling, the main features of Coulomb blockade and even can capture Kondo physics for temperatures of the order of or larger than the Kondo temperature~\cite{Koenig96prb}. The QME technique leads to more simple ``classical'' master equations in the case where (i) the electrode-system coupling can be considered as a weak perturbation, and (ii) off-diagonal elements of the reduced density matrix in the eigenstate representation (coherences) can be neglected due to very short decoherence times.

\index{Nonequilibrium Green functions !}
{\bf Nonequilibrium Green function (NGF)}~\cite{Kadanoff62book,Keldysh64,Langreth76inbook,Rammer86rmp,Haug96book} formalism is able to deal with a very broad variety of physical problems related to quantum transport.

The method, which was proposed about 50 years ago \cite{Kadanoff62book,Keldysh64}, then was used successfully in the theory of nonequilibrium superconductivity \cite{Eliashberg71,Larkin75a,Gray81,Langenberg86,Kopnin01book}, and later was proposed as a standard approach in mesoscopic physics and molecular electronics \cite{Meir92prl,Runge92prb,Jauho94prb}. We consider the general NGF formalism in chapter \ref{chapter:NGF} and the NGF method for transport through nanosystems in chapter \ref{MW}. The advantage of the NGF formalism is that it can be successfully applied to a variety of systems and problems, that it is in principle exact, and many powerful approximations can be derived from it.

It can deal with strong non-equilibrium situations via an extension of the conventional Green Function formalism to the Schwinger-Keldysh contour~\cite{Keldysh64} and it can also include interaction effects (electron-electron, electron-vibron, etc) in a systematic way (diagrammatic perturbation theory, equation of motion techniques). Proposed first time for the mesoscopic structures in the early seventies by Caroli et al.~\cite{Caroli71jpcss1,Caroli71jpcss2,Caroli71jpcss3,Caroli71jpcss4}, this approach was formulated in an elegant way by Meir, Wingreen and Jauho~\cite{Meir92prl,Wingreen93prb,Jauho94prb,Haug96book,Jauho06jpcs}, who derived exact expression for nonequilibrium current through an interacting nanosystem placed between large noninteracting leads in terms of the nonequilibrium Green functions of the nanosystem. Still, the problem of calculation of these Green functions is not trivial. We consider some possible approaches in the case of electron-electron and electron-vibron interactions. Moreover, as we will show later on, it can reproduce results obtained within the master equation approach in the weak coupling limit to the electrodes (Coulomb blockade), but it can also go {\it beyond} this limit and cover intermediate coupling (Kondo effect) and strong coupling (Fabry-Perot) domains. It thus offer the possibility of dealing with different physical regimes in a unified way.

Both approaches, the QME and NGF techniques, can yield formally exact expressions for many observables. For noninteracting systems, one can even solve analytically many models. However, once interactions are introduced - and these are the most interesting cases containing a very rich physics - different approximation schemes have to be introduced to make the problems tractable. We consider some examples of nonequilibrium problems in chapter \ref{chapter:MB}.

\index{Coulomb blockade !}
\textbf{Coulomb interaction} is in the origin of such fundamental effects as Coulomb blockade and Kondo effect. The most convenient and simple enough is the Hubbard-Anderson model, combining the formulations of Anderson impurity model~\cite{Anderson61prev} and Hubbard many-body model \cite{Hubbard63procroysoc,Hubbard64procroysoc1,Hubbard64procroysoc2}. To analyze strongly correlated systems several complementary methods can be used: master equation and perturbation in tunneling, equation-of-motion method, self-consistent Green functions, renormalization group and different numerical methods.

When the coupling to the leads is weak, the electron-electron interaction results in Coulomb blockade, the sequential tunneling is described by the master equation method~\cite{Averin86jltp,Averin91inbook,Grabert92book,vanHouten92inbook,Kouwenhoven97inbook,Schoeller97inbook,Schoen97inbook,vanderWiel02rmp} and small cotunneling current in the blockaded regime can be calculated by the next-order perturbation theory~\cite{Averin89pla,Averin90prl,Averin92inbook,Aleiner02pr}. This theory was used successfully to describe electron tunneling via discrete quantum states in quantum dots~\cite{Averin91prb,Beenakker91prb,vonDelft01pr,Bonet02prb}.  Recently there were several attempts to apply master equation to multi-level models of molecules, in particular describing benzene rings~\cite{Hettler03prl,Muralidharan06prb,Begemann08prb,Ryndyk13prb}.

To describe consistently cotunneling, level broadening and higher-order (in tunneling) processes, more sophisticated methods to calculate the reduced density matrix were developed, based on the Liouville - von Neumann equation~\cite{Petrov04chemphys,Petrov05chemphys,Petrov06prb,Elste05prb,Harbola06prb,Pedersen07prb,Mayrhofer07epjb,Begemann08prb} or real-time diagrammatic technique~\cite{Schoeller94prb,Koenig96prl,Koenig96prb,Koenig97prl,Koenig98prb,Thielmann03prb,Thielmann05prl,Aghassi06prb}.  Different approaches were reviewed recently in Ref.\,\cite{Timm08prb}.

The equation-of-motion (EOM) method is one of the basic and powerful ways to find the Green functions of interacting quantum systems. In spite of its simplicity it gives the appropriate results for strongly correlated nanosystems, describing qualitatively and in some cases quantitatively such important transport phenomena as Coulomb blockade and Kondo effect in quantum dots. The results of the EOM method could be calibrated with other available calculations, such as the master equation approach in the case of weak coupling to the leads, and the perturbation theory in the case of strong coupling to the leads and weak electron-electron interaction.

In the case of a single site junction with two (spin-up and spin-down) states and Coulomb interaction between these states (Anderson impurity model), the \emph{linear conductance} properties have been successfully studied by means of the EOM approach in the cases related to Coulomb blockade\cite{Lacroix81jphysf,Meir91prl} and the Kondo effect~\cite{Meir93prl}. Later the same method was applied to some two-site models~\cite{Niu95prb,Pals96jpcm,Lamba00prb,Song07prb}. Multi-level systems were started to be considered only recently~\cite{Palacios97prb,Yi02prb}. Besides, there are some difficulties in building the lesser GF in the nonequilibrium case (at finite bias voltages) by means of the EOM method~\cite{Niu99jpcm,Swirkowicz03prb,Bulka04prb}.

The diagrammatic method was also used to analyze the Anderson impurity model. First of all, the perturbation theory can be used to describe weak electron-electron interaction and even some features of the Kondo effect~\cite{Fujii03prb}. The family of nonperturbative current-conserving self-consistent approximations for Green functions has a long history and goes back to the Schwinger functional derivative technique, Kadanoff-Baym approximations and Hedin equations in the equilibrium many-body theory~\cite{Schwinger51pnas,Martin59prev,Baym61prev,Baym62prev,Hedin65prev,Gunnarsson88rpp,White92prb,Onida02rmp}.  Recently GW approximation was investigated together with other methods~\cite{Thygesen07jcp,Thygesen08prl,Thygesen08prb,Wang08prb}. It was shown that dynamical correlation effects and self-consistency can be very important at finite bias.

\index{Vibrons and polarons !}
{\bf Vibrons} (the localized phonons) is the other important ingredient of the models, describing single molecules because molecules are flexible. The theory of electron-vibron interaction has a long history, but many questions it implies are not answered up to now. While the isolated electron-vibron model can be solved exactly by the so-called polaron or Lang-Firsov transformation \cite{Lang63jetp,Hewson74jjap,Mahan90book}, the coupling to the leads produces a true many-body problem. The inelastic resonant tunneling of {\em single} electrons through the localized state coupled to phonons was considered in Refs.~\cite{Glazman88jetp,Wingreen88prl,Wingreen89prb,Jonson89prb,Cizek04prb,Cizek05czechjp}.  There the exact solution in the single-particle approximation was derived, ignoring completely the Fermi sea in the leads. At strong electron-vibron interaction and weak couplings to the leads the satellites of the main resonant peak are formed in the spectral function.

The essential progress in calculation of transport properties in the strong electron-vibron interaction limit has been made with the help of the master equation approach~\cite{Braig03prb,Aji03condmat,Mitra04prb,Koch05prl,Koch06prb,Koch06prb2,Wegewijs05condmat,Zazunov06prb1,Ryndyk08prb,Ryndyk10prb}.  This method, however, is valid only in the limit of very weak molecule-to-lead coupling and neglects all spectral effects, which are the most important at finite coupling to the leads.

At strong coupling to the leads and the finite level width the master equation approach can no longer be used, and we apply alternatively the nonequilibrium Green function technique which have been recently developed to treat vibronic effects in a perturbative or self-consistent way in the cases of weak and intermediate electron-vibron interaction~\cite{Tikhodeev01surfsci,Mii02surfsci,Mii03prb,Tikhodeev04prb,Galperin04nanolett,Galperin04jcp,Galperin05jphyschemb,Frederiksen04master,Frederiksen04prl,Hartung04master,Ryndyk05prb,Ryndyk06prb,Ryndyk07prb,Paulsson05prb,Paulsson06jpconf,Arseyev05jetplett,Arseyev06jetplett,Zazunov06prb}.

The case of intermediate and strong electron-vibron interaction {\em at intermediate coupling to the leads} is the most interesting, but also the most difficult. The existing approaches are mean-field~\cite{Hewson79jphysc,Galperin05nanolett,DAmico08njp}, or start from the exact solution for the isolated system and then treat tunneling as a perturbation~\cite{Kral97,Lundin02prb,Zhu03prb,Alexandrov03prb,Flensberg03prb,Galperin06prb,Zazunov07prb}.  The fluctuations beyond mean-field approximations were considered in Refs.\,~\cite{Mitra05prl,Mozyrsky06prb}

In parallel, the related activity was in the field of single-electron shuttles and quantum shuttles~\cite{Gorelik98prl,Boese01epl,Fedorets02epl,Fedorets03prb,Fedorets04prl,McCarthy03prb,Novotny03prl,Novotny04prl,Blanter04prl,Smirnov04prb,Chtchelkatchev04prb}.  Finally, based on the Bardeen's tunneling Hamiltonian method \cite{Bardeen61prl,Harrison61pr,Cohen62prl,Prange63pr,Duke69book} and Tersoff-Hamann approach~\cite{Tersoff83prl,Tersoff85prb}, the theory of inelastic electron tunneling spectroscopy (IETS) was developed~\cite{Persson87prl,Gata93prb,Tikhodeev01surfsci,Mii02surfsci,Mii03prb,Tikhodeev04prb,Raza08prb}.

The recent review of the electron-vibron problem and its relation to the molecular transport see in Ref.~\cite{Galperin07jpcm}.

Finally, we want to mention briefly three important fields of research, that we do not consider in this book: the theory of Kondo effect~\cite{Glazman88jetplett,Ng88prl,Hershfield92prb,Meir93prl,Wingreen94prb,Aguado00prl,Kim01prb,Plihal01prb}, spin-dependent transport~\cite{Sanvito06jctn,Rocha06prb,Naber07jpdap,Ke08prl,Ning08prl}, and time-dependent transport~\cite{Jauho94prb,Grifoni98pr,Kohler05pr,Sanchez06prb,Maciejko06prb}. 